# Simulating a guitar with a conventional sonometer

**Zily Burstein, Christina M. Gower and Gabriele U. Varieschi,** Loyola Marymount University, Los Angeles, CA 90045

Musical acoustics is an interesting sub-field of physics which is usually able to engage students in a dual perspective, by combining science and art together. The physics principles involved in most musical instruments[1] can be easily demonstrated with standard laboratory equipment and can become part of lecture or lab activities. In particular, we will show in this paper how to simulate a guitar using a conventional sonometer, in relation to the problem of the instrument intonation, i.e., how to obtain correctly tuned notes on a guitar or similar string instruments.

This problem is more complex than what might appear at first; it obviously begins with the correct tuning of the "open" strings of the instrument to the desired notes, which can be easily accomplished with the help of a digital tuner. It is then related to the correct placement of the "frets" on the "fingerboard" which enable the instrument to produce all the different notes, but it is further complicated by other subtle effects, which are usually dealt with in an empirical way by luthiers and guitar manufacturers.

In a recently published paper[2] we described mathematical and physical models to be used for a more scientific approach to this problem, resulting in a complex "compensation" procedure which is very effective in improving the intonation of this type of instruments. In this paper we present a simplified approach to the problem, which is more suitable to be used as an in-class demonstration, or as a laboratory activity, to simulate how musical notes are produced on a guitar.

### The experimental apparatus and the intonation problem

A conventional (monochord) sonometer, such as the PASCO[3] WA-9613 or similar, can be used to simulate a (single string) guitar, with the addition of a fingerboard which can be easily obtained from a local luthier or through a music shop. Figure 1 shows our experimental apparatus composed of the PASCO sonometer to which we added part of a classical guitar fingerboard, complete with twenty metallic frets, placed right under the steel string.

It is important to know the scale length of the fingerboard (or fretboard) being used, i.e., the correct string length for which the fingerboard was designed (usually between 640-660 mm for a classical guitar) and to set the distance between the two sonometer bridges accordingly. These two bridges will simulate the "nut" and the "saddle" of a guitar, as shown in Fig. 1. A "guitar outline" is also superimposed on this figure, to illustrate how our apparatus can simulate the functionality of a real guitar.

The frets are placed according to a precise mathematical relation:

$$X_i = X_0 2^{-\frac{i}{12}} \cong X_0 (0.943874)^i, \qquad (1)$$

where $X_0$ is the (open) string length, or scale length, mentioned above and $X_i$ is the position of the i-th fret, as measured from the saddle of the instrument (see our previous paper[2] for full details).

The fretboard needs to be positioned right under the sonometer string so that the 12th fret is placed exactly at half the string length (as required by Eq. (1) for $i = 12$) and as close as possible to the string as in a real guitar. After tuning the open string to the desired note, one should test all the other notes by pressing down on each fret and plucking the string. Usually the sonometer has enough resonance to produce an audible sound, although not as strong as the one of a real instrument.

In this setup process, care should be taken to adjust the "action" of our simulated instrument, i.e., placing the fretboard close enough to the string so that it is easy to press down on it at any position along the fingerboard, but not too close in order to avoid producing "buzzing" sounds when playing the notes on the fretboard. A good compromise can usually be achieved after trying different heights of the fretboard above the base level of the sonometer. To press uniformly on the string at any position we used a spring-loaded device (also shown in Fig. 1), but finger pressure will also work. The string can be plucked by hand or with a standard guitar pick.

After this initial setup a chromatic scale can be played, i.e., all the notes on the fretboard in succession, and the frequency of the produced sounds can be measured, to check if the simulated guitar is in tune. To measure these frequencies one can use a microphone connected to a digital oscilloscope or to a digital interface linked to a computer, using one of the many software packages commercially available, or other devices. In our previous work[2] we preferred to use a very precise digital tuner[4] which could discriminate frequencies at the level of $\pm 1$ cent[5].

The measured frequencies should be checked against the theoretical frequencies which are expressed in a way similar to Eq. (1):

$$\nu_i = \nu_0 2^{\frac{i}{12}} \cong \nu_0 (1.05946)^i, \qquad (2)$$

where $\nu_0$ and $\nu_i$ are respectively the frequency of the open string note and of the i-th note. Eqs. (1) and (2) are simply related, since Mersenne's law[2] states that the frequency of a vibrating string is inversely proportional to its length, and the vibrating length of a guitar string is expressed by $X_i$ in Eq. (1).

At this point, a trained ear should perceive that the sonometer is not perfectly in-tune, even if the setup procedure outlined above was perfectly followed. This lack of intonation, noticeable for most of the "fretted" notes, i.e., those obtained by pressing the string onto the fretboard, is a well-known effect in guitar manufacturing.

It is mainly due to the mechanical action of the player's finger, which presses the string on the fingerboard, thus changing slightly its length and tension and therefore altering the frequency of the sound being produced. As a result, a "fretted" string instrument, such as a guitar, mandolin, or similar instrument, is never perfectly in tune, since the pressure of the player's finger is unavoidable. On the contrary, "non-fretted" string instruments, such as those of the violin family, do not have this complication, as long as a skilled player can compensate for this effect by pressing on the fingerboard at a position which naturally eliminates the problem.

Figure 2 shows a detail of our apparatus, illustrating how the spring-loaded device is pressing at a particular position (the fifth fret in this case), thus altering the original length and tension of the string, in the same way the player's finger acts on a real guitar fingerboard. The digital tuner, placed near the sonometer, can accurately measure the frequency of the produced sound, when the string is plucked.

In the next section we will outline how a particular compensation procedure can minimize this lack of intonation of a fretted instrument and show how it can be practically implemented on our monochord sonometer.

**The compensation procedure**

In order to correct the intonation problem described above the compensation procedure usually involves moving slightly both saddle and nut, from their initial positions. In particular, the saddle is usually moved a few millimeters away from the nut ("setback" of the saddle, increasing the open string length), while the nut is usually moved forward toward the saddle by a smaller amount ("set-forth" of the nut, decreasing the open string length). These movements are performed without changing the positions of all the frets and they can be easily performed, in our experimental apparatus, by adjusting the positions of the two movable bridges of the sonometer which simulate the saddle and the nut.

In our previous work[2] on the subject, we analyzed a complex mathematical procedure which maximized the compensation effects, and we computed the best possible saddle-nut positions. In this paper we describe a simplified procedure, similar to the one used by guitar manufacturers, which can be effectively used as a lab activity for students in a musical acoustics class.

The procedure we used involves the following steps:

1. Set the nut and saddle at their initial positions for the string length being used and tune the string to the required pitch (open string note).
2. Using a spring-loaded device or a finger, press at all fret positions and record the frequencies of all notes *without compensation* (at least three measurements for each fret, in order to compute average and standard deviation for each frequency).
3. Move the nut to a new set-forth position (by small fixed increments, of one millimeter or less) and retune the open string to its proper value.
4. Compare the fretted note at the 5$^{th}$ fret with the theoretical frequency expected for that note (obtained using Eq. (2)).
5. Return to point 3 and iterate the procedure until the best position for the nut is found (the one for which the fretted 5$^{th}$ note is more in-tune). Record the final nut set-forth.
6. Leaving the nut at this new position, proceed to adjust the saddle to a new setback position (by small fixed increments, of one millimeter or less for example) and then retune the open string to its proper value.
7. Compare the fretted note at the 12$^{th}$ fret with the theoretical frequency (obtained using Eq. (2)).
8. Return to point 6 and iterate the procedure until the best position for the saddle is found (the one for which the fretted 12$^{th}$ note and the corresponding theoretical frequency are best matched). Record the final saddle setback.
9. Leave the saddle and nut at these new compensated positions and retune once again the open string to the proper frequency.

10. Finally, press at all fret positions and record the frequencies of all notes *with compensation* (three measurements for each fret, average and standard deviation for each frequency).

**Experimental results**

We applied the procedure outlined above to two steel strings included in the PASCO sonometer and also used in our past analysis[6] (see Table I in our previous paper[2] for the physical properties of these strings).

The first string was tuned at an open string frequency $v_0 = 130.813$ Hz, corresponding to a $C_3$ note. Using Eq. (2), it is straightforward to compute the theoretical frequencies of all the notes corresponding to the twenty frets of our simulated guitar. We then used the procedure described above to measure all the experimental frequencies of the notes with and without compensation, and compared these values with the theoretical ones.

The results are more effectively presented in terms of the frequency deviation (in cents[5]) between the experimental and the theoretical values. A zero frequency deviation for a particular note means perfect tuning, while a frequency deviation of about $\pm 10$ cents (pitch discrimination range) would still be considered acceptable, since the human ear cannot differentiate two sound frequencies within this range (see Ref. 2 for more details).

Therefore, in Fig. 3 the frequency deviations for the first string are presented as a function of the fret number of our simulated guitar. The perfect intonation level corresponds to zero frequency deviation (black dotted line) and it is obviously achieved only by the open string (zero fret number). Red circles represent results without compensation, while blue triangles are obtained with our compensation procedure (with a nut set-forth of 5.0 mm and a saddle setback of 5.0 mm). Results within the pitch discrimination range (green dashed lines) should be considered practically in-tune, in view of the previous considerations. We can see that the compensation procedure was effective since the results with compensation are in fact within the required range. The nut adjustment corrected the intonation at lower fret numbers, while the saddle adjustment took care of the higher frequency notes.

Fig. 4 presents similar results for a second string, which was tuned an octave higher than the first one (open string frequency $v_0 = 261.626$ Hz, corresponding to a $C_4$ note). The meaning of the symbols in this figure is the same as in Fig. 3, and again we can see that the compensation procedure was effective (with a nut set-forth of 3.0 mm and a saddle setback of 2.0 mm for this string).

**Conclusions**

We presented a simple way to convert a standard sonometer into a simulated fret instrument, by adding a guitar fingerboard. This modified apparatus can be useful to introduce students to string vibrations, Mersenne's law and its applications to fretted string instruments. The subtle effects of intonation and compensation can also be studied in a practical way, without using complex mathematical models. The experiments presented in this work can easily become a more structured laboratory activity, which can be used in general physics courses or more specialized acoustics classes.

**Acknowledgments**

This research was supported by a grant from the Frank R. Seaver College of Science and Engineering, Loyola Marymount University. The authors would like to acknowledge luthiers Greg Byers and Hugh Greenwood for useful discussions and advice.

**Zily Burstein** is a sophomore honors student at Loyola Marymount University, majoring in physics and English. She is interested in exploring many areas of physics during her remaining three years at LMU, before attending graduate school.
**Department of Physics, Loyola Marymount University, 1 LMU Drive,
Los Angeles, CA 90045; zburstei@lion.lmu.edu**

**Christina M. Gower** is a senior physics and mathematics major at Loyola Marymount University. During her four years at LMU Christina has conducted research in different areas of physics including space physics, condensed matter, acoustics, and cosmology. Christina is planning to attend graduate school in astrophysics in the fall of 2011.
**Department of Physics, Loyola Marymount University, 1 LMU Drive,
Los Angeles, CA 90045; Christina.Gower@gmail.com**

**Gabriele U. Varieschi** is an associate professor of physics at Loyola Marymount University. He received his Ph.D. in theoretical particle physics from the University of California at Los Angeles. His research interests are in the areas of astro-particle physics, cosmology and general physics. In his free time he enjoys playing classical guitar and renaissance lute music.
**Department of Physics, Loyola Marymount University, 1 LMU Drive,
Los Angeles, CA 90045; gvarieschi@lmu.edu**

**FIGURES:**

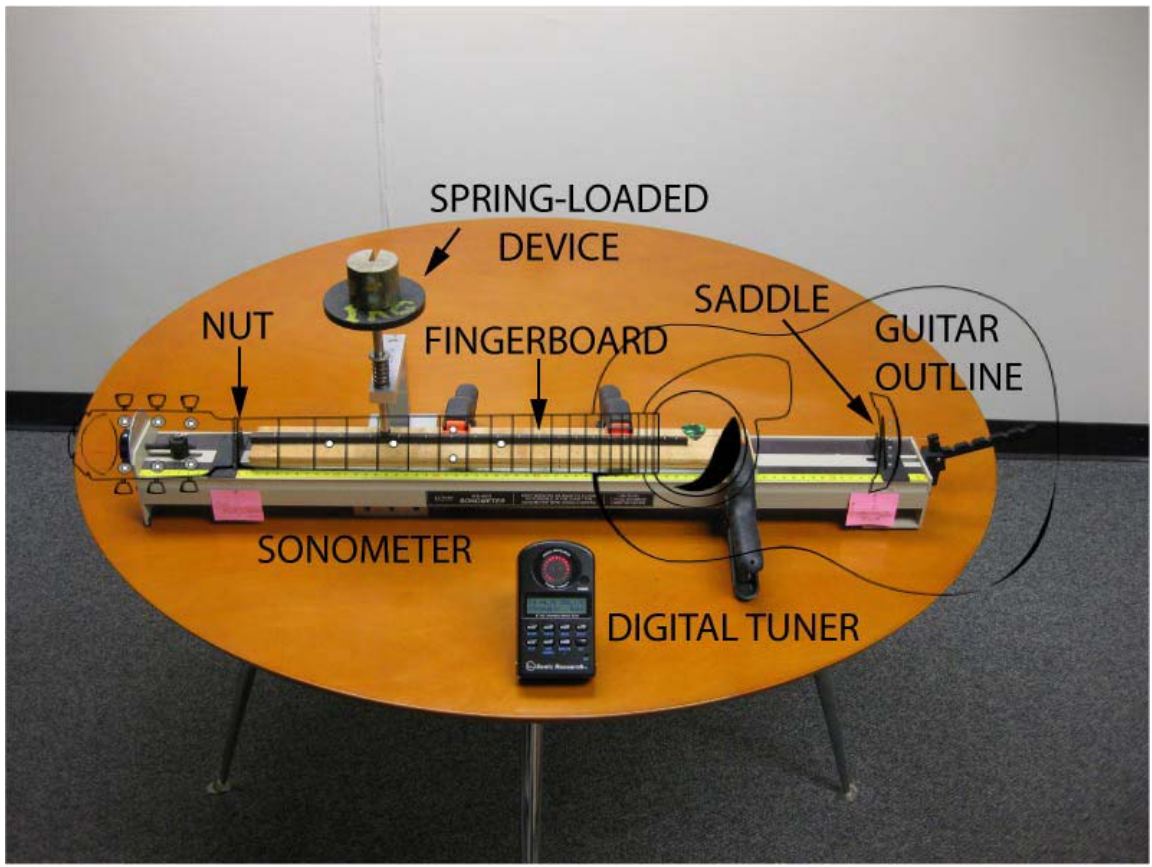

Fig. 1. Our tabletop apparatus composed of a sonometer and a guitar fingerboard, with twenty metallic frets. Also shown are the two movable bridges, which simulate the nut and saddle of a guitar, the spring-loaded device used to press on the string at different positions, and the digital tuner used to measure sound frequencies. A guitar outline is superimposed on the picture, in order to show how the apparatus simulates the functionality of a real guitar.

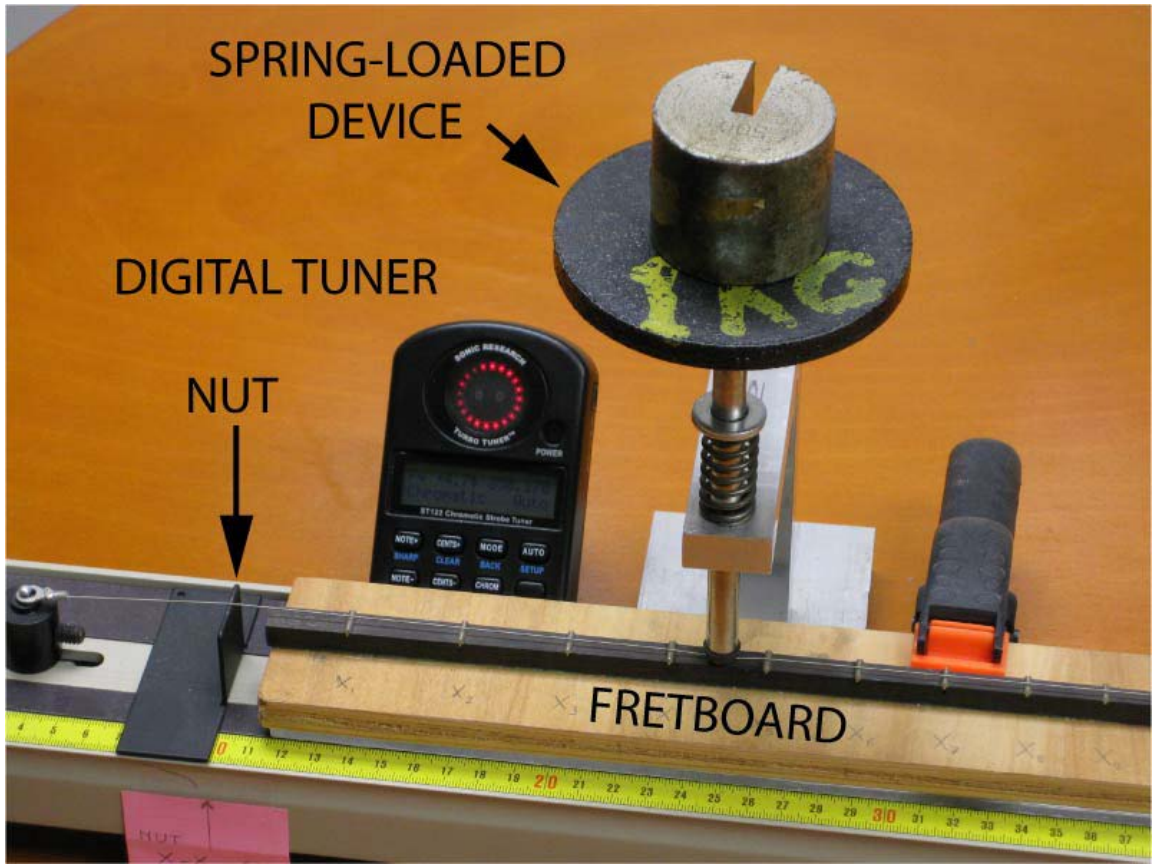

Fig. 2. A detail of our experimental apparatus, showing the spring-loaded device pressing at the fifth fret position. The resulting string deformation and the slight change in the string tension will cause the lack in intonation of the note being produced. The digital tuner positioned near the sonometer can accurately measure the sound frequency.

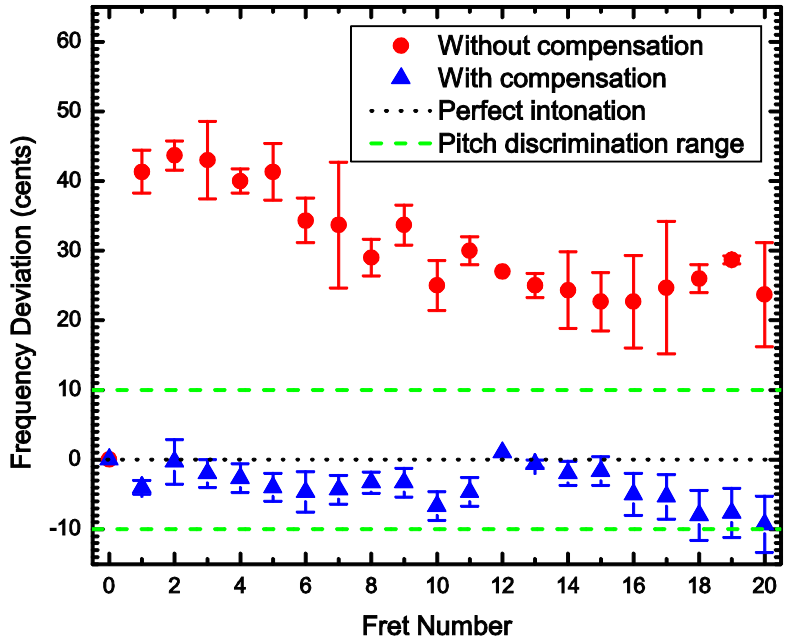

Fig. 3. Frequency deviation from perfect intonation level for notes obtained with our first string. Red circles indicate results without compensation, while blue triangles denote results with compensation. The region between the green dashed lines is the approximate pitch discrimination range for frequencies related to this string.

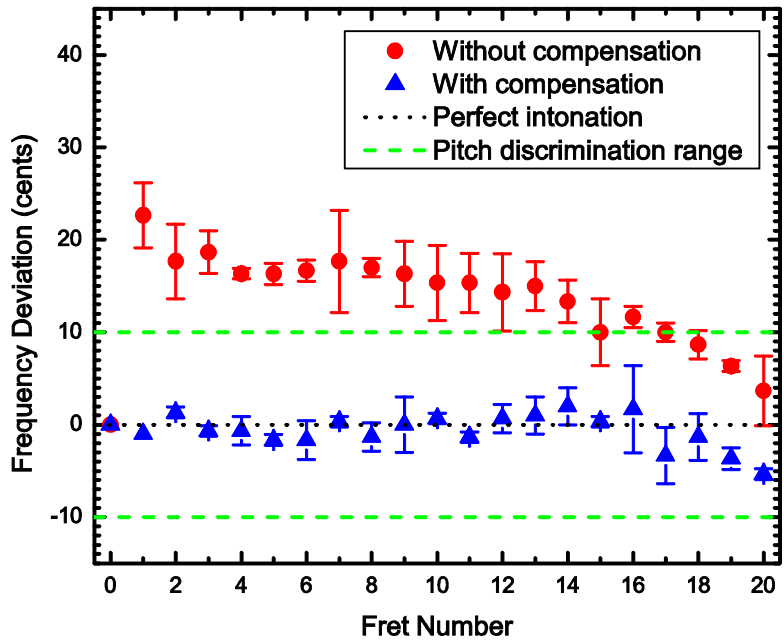

Fig. 4. Frequency deviation from perfect intonation level for notes obtained with our second string. Red circles indicate results without compensation, while blue triangles denote results with compensation. The region between the green dashed lines is the approximate pitch discrimination range for frequencies related to this string.